\documentclass[reprint,amsmath,amssymb,aps,pra,floatfix]{revtex4-2}
\usepackage[utf8]{inputenc}
\usepackage{graphicx}
\usepackage{xcolor}
\usepackage{braket}
\usepackage{physics}
\usepackage{upgreek}
\usepackage{xcolor}
\usepackage{url}
\usepackage{textcomp}
\usepackage{xr-hyper}
\usepackage[title]{appendix}
\usepackage[colorlinks]{hyperref}

\begin{document}

\title{Design and implementation of a modular laser system for AMO experiments}

\author{Klara Theophilo, Scott J Thomas, Georgina Croft, Yashna N D Lekhai, Alexander Owens, Daisy R H Smith, Silpa Muralidharan and Cameron Deans}
\email{cameron.deans@stfc.ac.uk}

\affiliation{National Quantum Computing Centre, Rutherford Appleton Laboratory, OX11 0QX, UK}
\date{\today}

\begin{abstract}
    Robust laser delivery and stabilization are key components in atom-based quantum technologies, such as quantum computing. Moving these technologies towards product-like deployment requires scalable, compact, cost-effective, and upgradable modules. Here we describe laser systems consisting of application-flexible modules and demonstrate their performance by characterizing key metrics and by integration with ion trap systems. The laser system is confined to a single server rack and a compact locking station. Both are Class~1 laser products with fiber in-out and electronic control of the laser light. This is achieved through precision manufacture of optical boards that are designed to reduce the degrees of freedom, ease alignment, and increase the robustness to environmental factors. We present a range of 13 wavelengths from 375~nm to 1092~nm: efficiencies from laser source to ion trap range from 21 - 28\%, with laser stabilization line widths below $1$~MHz. 
\end{abstract}
\maketitle

\section{Introduction}

Laser trapping, cooling, and coherent population pumping are key pillars of atomic physics platforms \cite{Metcalf1999Laser}. Achieving coherent control typically imposes strict requirements on the laser light and consequently, the laser system. Alongside the laser performance parameters, it is becoming increasingly important to consider requirements that enable the applications of these laser systems beyond a laboratory environment, such as: system footprint, portability, reproducibility, modularity, and ease-of-use. For instance, scaling quantum computers towards fault tolerance requires thousands of qubits, and therefore footprint and reproducibility of the quantum processing units (QPUs) become as relevant as their efficiency \cite{osti_2588210, ConnectIons, Data0,quantumData1}.

Here we describe a reproducible, modular, high-performance laser system implemented across a wide range of wavelengths. The systems reported here are used for trapped-ion quantum computing \cite{PhysRevLett.74.4091, 10.1063/1.5088164, Hempel2026}; however, they are adaptable to technologies with varied requirements. The core principle is the development of custom optical boards that significantly reduce of the number of components and alignment degrees of freedom. This approach shifts the setup burden away from the end user to the design and manufacture of the boards. The result is an optical setup that is significantly faster to assemble and align, cheaper, and more compact than typical table-top implementations. Laser safety is built in from concept with the majority of the optical boards housed in a fully enclosed 19-inch rack. Overall, our system aims to shift the requirement for laser light that is fully controllable in power, frequency, and polarization from a time-consuming combination of designing, building, and maintaining to a stand-alone trusted and tested product. 

\begin{figure}[!ht]
\includegraphics[width=0.9\linewidth]{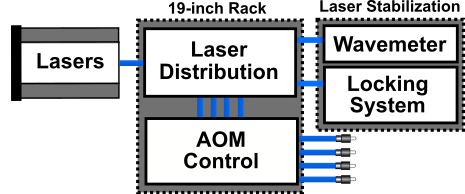}
\caption{Schematic of the laser system. Optical fibers connect the laser sources (see supplementary material Sec.~\ref{sec:Laser_sources}) to the distribution modules. Each distribution module has six flexible outputs. In our implementations, one goes to a wavemeter, one goes to a locking system, and the remainder go to AOM modules -- controlling up to four experimental outputs. In total we fit 14 optical boards into a 39~U 19-inch rack -- each distribution optical board contains two distribution modules and each AOM board contains four AOM modules.}
\label{fig:LaserSk}
\end{figure}

A schematic of the modular laser system is presented in Fig. \ref{fig:LaserSk}. It comprises three strands, each connected via optical fibers: laser sources, laser distribution and control (housed in a 19-inch standard server rack), and laser stabilization (connecting a wavemeter and locking solution). Each wavelength is distributed into 6 variable power outputs, that are fed either to the stabilization module or into control modules featuring acousto-optical modulators (AOMs) that enable frequency shifting and pulse control. 

To evaluate the system we measure and discuss the following performance metrics: overall power/efficiency and power stability, frequency stability, and polarization stability. These requirements are ubiquitous in atomic physics. For example, for cooling and optical pumping, the frequency must be stabilized to a value smaller than the natural linewidth $\Gamma/2\pi$ atomic transition  being addressed (typically in the MHz range \cite{Steck2007}, for example 20.7~MHz for $^{40}\text{Ca}^+$ \cite{Poulsen11} and 20.4~MHz for $^{88}\text{Sr}^+$ \cite{Valli2004}). Moreover, to address different transitions or perform optical molasses \cite{Metcalf1999Laser}, the frequency must also be scannable. Equally, laser power ($P$) and power stability have a direct impact on coherent operations, given atomic excitation probability is proportional to the Rabi frequency ($\Omega$), and $\Omega \propto \sqrt{P}$ \cite{allen2012optical}. Furthermore, for transitions using Zeeman states, polarization needs to be well defined and maintained to ensure correct addressing \cite{cohen2019quantum}.

A detailed description of the design concept and each of the modules is presented in Sec.~\ref{sec:DC}, followed by the performance of the modules in Sec.~\ref{sec:results}. We present data from three separate instances of this laser system designed for two ion species (calcium and strontium) over a wide band of wavelengths: 375~nm, 395~nm, 397~nm, 422~nm, 423~nm, 461~nm (`blues'), 674~nm, 729~nm, 854~nm, 866~nm (`reds'), and 1004~nm, 1033~nm, 1092~nm (`infrareds').

\section{Design Concept}
\label{sec:DC}

\subsection{In-Rack Modules}
Traditional optics setups are hand-placed and clamped. This results in a high number of degrees of freedom. Minimizing the degrees of freedom not only increases precision and reproducibility, but reduces the time required for alignment. The number of degrees of freedom is reduced by approximately 70$\%$ compared to a typical table-top setup by employing machined boards and positioning via dowel pins. For example, our previous table-top double-pass AOM setup had 52 degrees of freedom, while our AOM module has 17. The boards minimize the footprint of the system with increased mechanical stability due to the reduction in components.

\subsubsection{Distribution Module}
\begin{figure}[!ht]
\includegraphics[width=8cm]{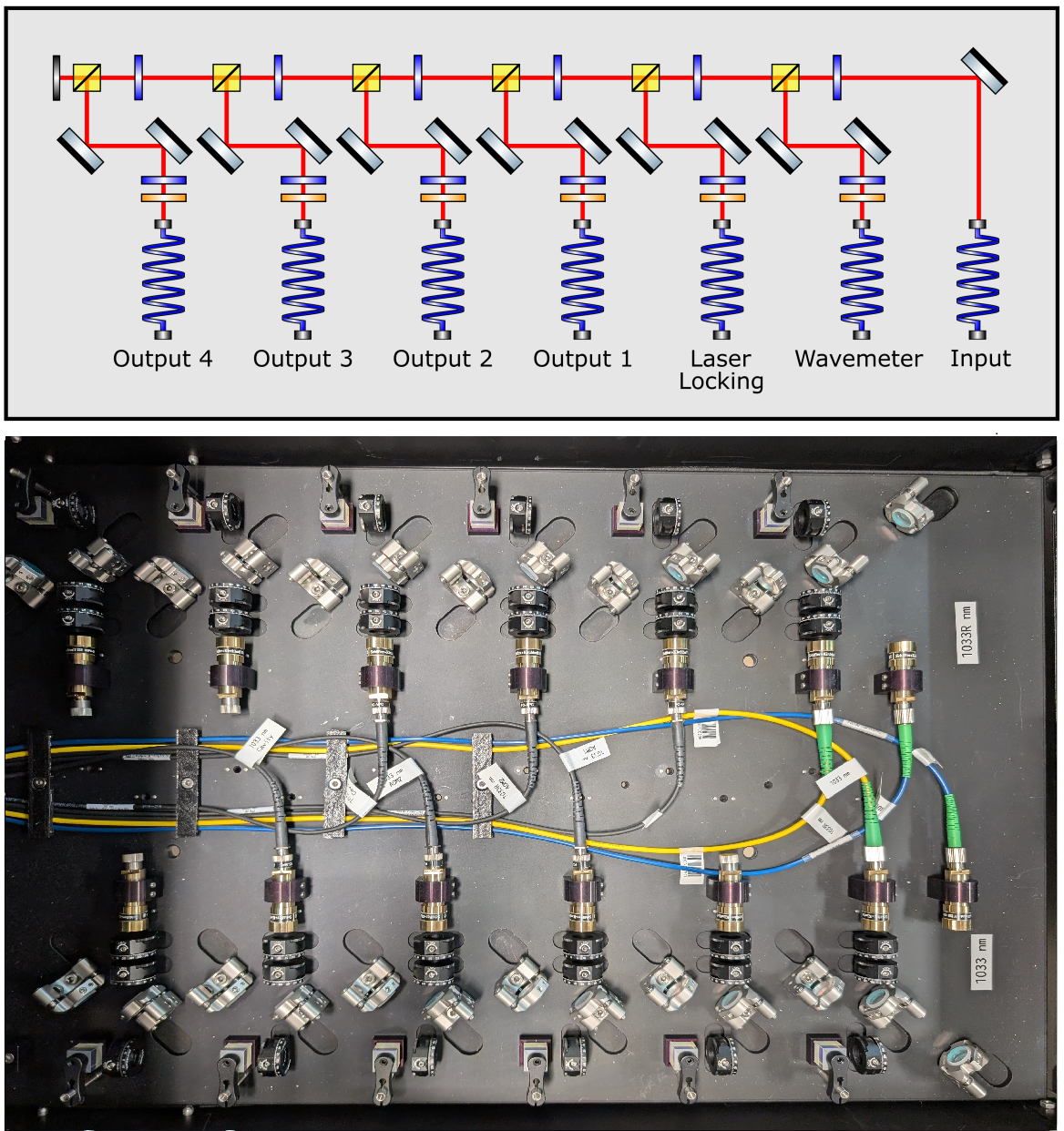}
\caption{(Top) Schematic of the distribution module (see Appendix~\ref{sec:Schematics} for the key). (Bottom) Photo of a completed distribution board. Each board consists of two distribution modules (for two separate wavelengths). Brackets are used for strain relief and routing of fibers.}
\label{fig:distributionschematic}
\end{figure}

The distribution module, shown in Fig. \ref{fig:distributionschematic}, takes a single input optical fiber and splits laser light into six variable power outputs. We use four outputs for experimental use and route each to an AOM module, whilst the other two provide frequency monitoring (wavemeter - High-Finesse WS8-10) and frequency stabilization. The power distribution in each output can be controlled by a combination of half-waveplate and polarizing beam splitter (PBS). Polarization on the output fiber is controlled with a quarter and a half-waveplate before the collimator. The input and output collimators are the same, to maximize fiber coupling efficiency.

\subsubsection{Acousto-optical Modulator (AOM) Modules}

\begin{figure}[!ht]
\includegraphics[width=8cm]{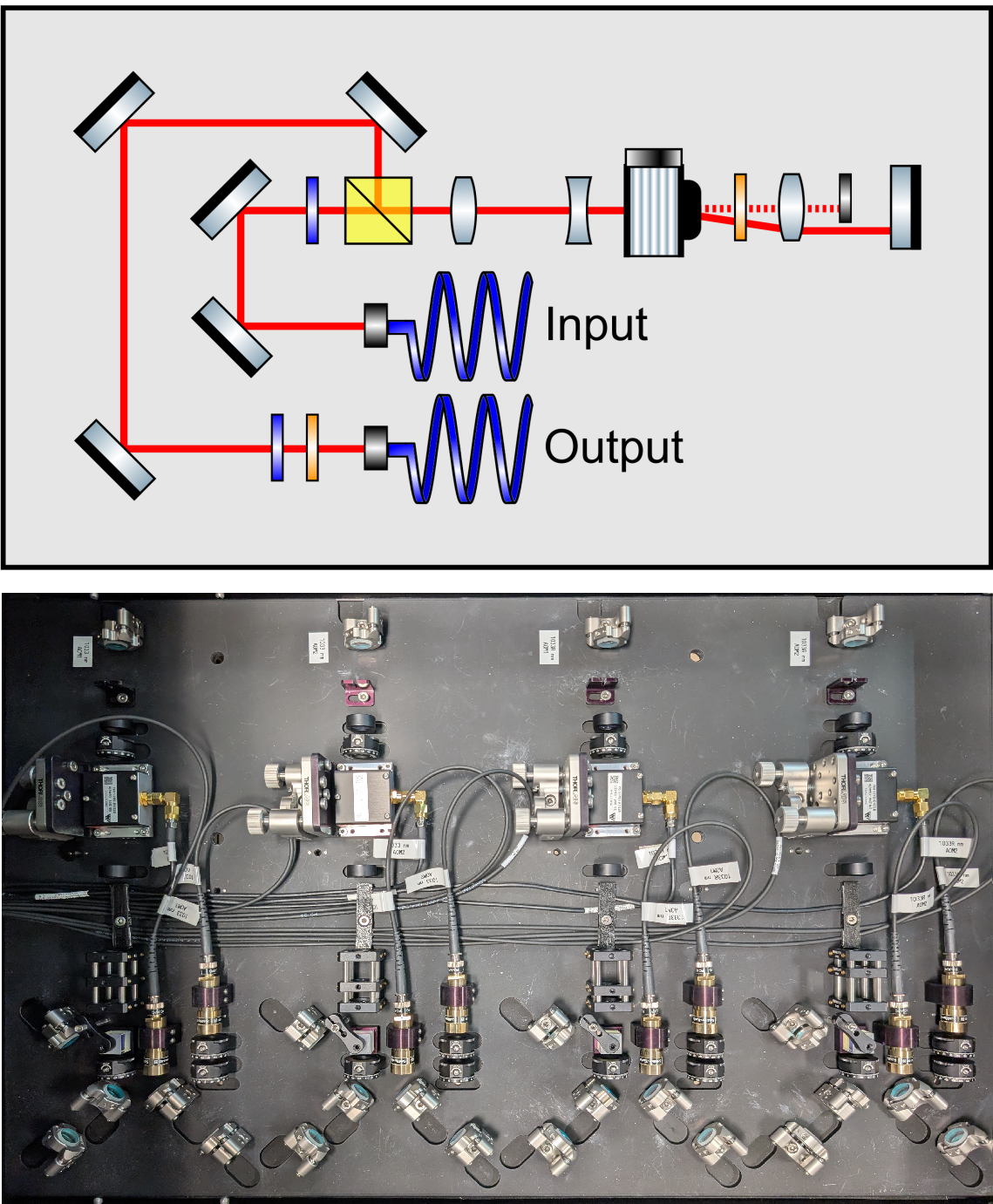}
\caption{(Top) Schematic of a single double-pass AOM module (see Appendix~\ref{sec:Schematics} for the key). (Bottom) Photo of a completed AOM board. Each board consists of four double-pass AOM modules -- each providing independent control.}
\label{fig:AOMBoards}
\end{figure}

Each unit allows for control of the frequency and amplitude of an individual beam. The AOMs are placed in a double-pass configuration \cite{Donley10.1063/1.1930095}. Two key design choices reduce the footprint of the module: a Galilean telescope to reduce beam waist at the AOM and a cat's eye reflector to minimize spatial displacement with changing frequency. Combined, these allow four AOM modules to be placed on one board (Fig. \ref{fig:AOMBoards}). The choice of lenses and beam size are determined by wavelength and experimental requirements (see Sec.~\ref{sec:AOMEff}). The AOMs featured here were chosen due to their large bandwidth.

\subsubsection{Dichroic Combination Module}

Combining laser beams can be achieved using PBSs, beam splitters, or dichroic mirrors. Beam splitters always incur a power loss equivalent to their splitting ratio (transmitted light:reflected light). PBSs have high transmission for both inputs, but require differing polarizations. Dichroic mirrors are therefore usually preferable, having no polarization requirement and high transmission for both inputs. Suitable elements are challenging for similar wavelengths -- with a separation $\geq 5$~nm typically required.

\begin{figure}[!ht]
\includegraphics[width=0.85\linewidth]{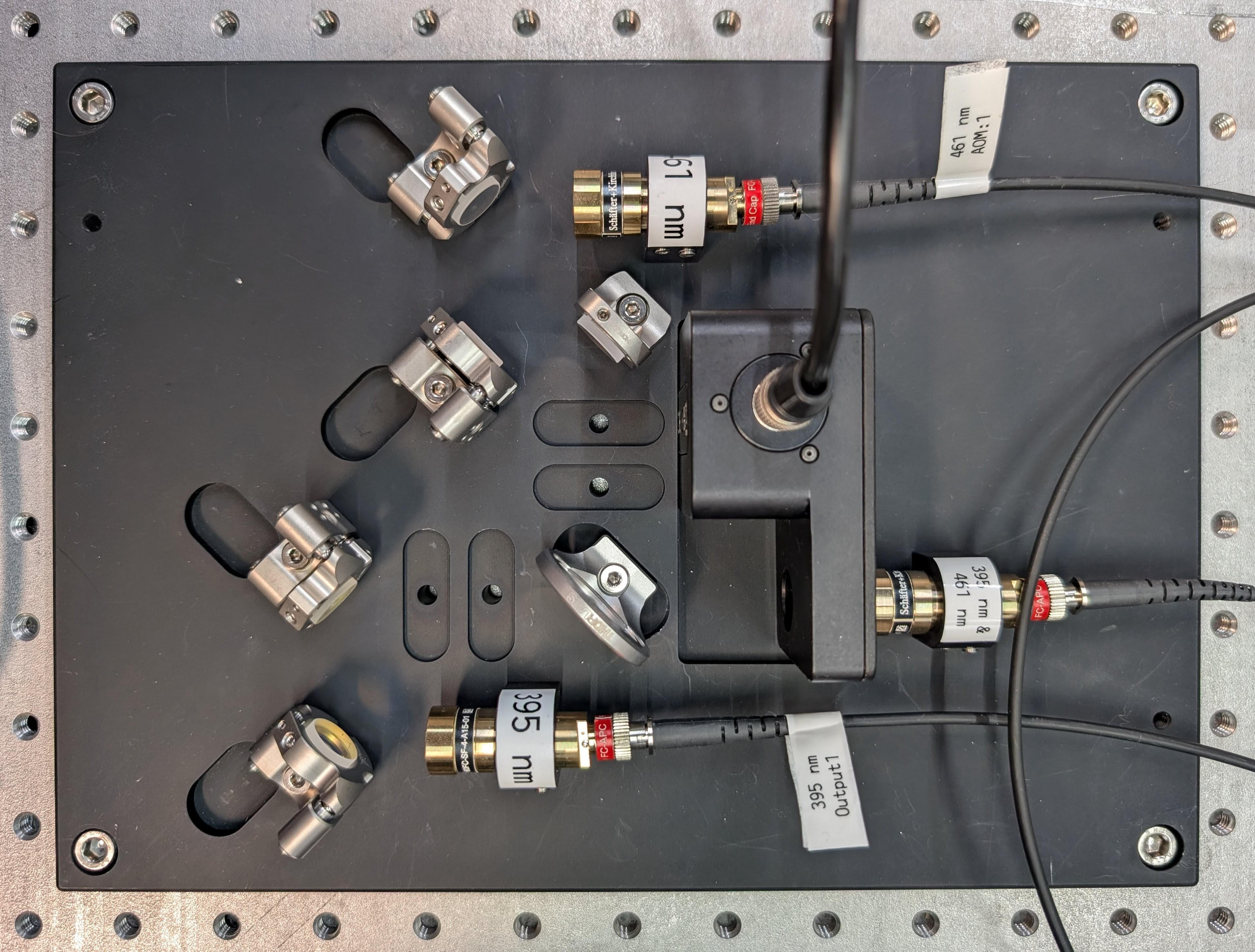}
\caption{Photo of a dichroic combination module for combining two wavelengths. Example shown: combining 395~nm and 461~nm provides the two-stage photoionisation beams for strontium ion trapping.}
\label{fig:combiboard}
\end{figure}

Our dichroic combination module employs a dichroic mirror to overlap a pair of laser beams and inject them into the same fiber. This simplifies alignment when multiple beams need to be delivered to individual atoms --  for example two-stage photoionisation \cite{PhotoIons}. An optional shutter can be included to control the light delivery. Fig.~\ref{fig:combiboard} shows an example dichroic combination module combining 395~nm and 461~nm light for two-stage photoionisation of strontium. 

\subsection{19-inch Rack Design and Drawers}

\begin{figure}[!ht]
\includegraphics[width=6cm]{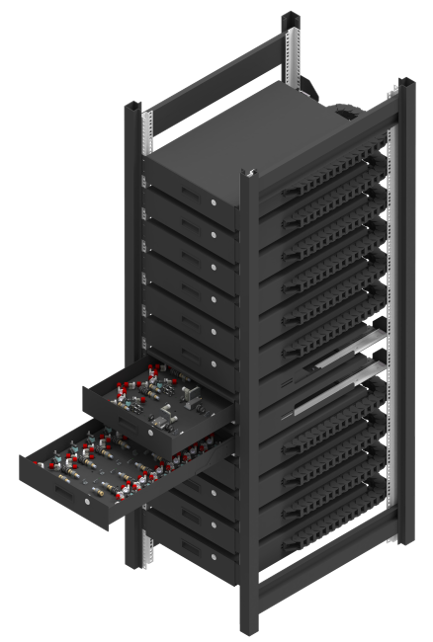}
\caption{Rendering of the complete rack-based laser system, with the optical boards housed inside light-tight drawers.}
\label{fig:rack_mounting}
\end{figure}

The modules are designed with custom-made drawers compatible with standard 19-inch racks (`server racks'). Factors influencing the design include laser safety, user access, optical fiber routing, and footprint. A rendering of a complete rack can be seen in Fig.~\ref{fig:rack_mounting}.

We use a 600~mm $\times$ 800~mm $\times$ 39~U rack, where U is a rack unit (equal to 44.45~mm and denotes the spacing between evenly placed mounting points). The height of each drawer module is 110~mm. This allows for 14 drawers within each rack. 5 distribution boards, 8 AOM boards, and 1 electro-optical modulator (EOM) board provides full experimental control for a minimum of 8 wavelengths, with 4 experimental outputs each. For many standard atomic species, this is sufficient for all required laser control, for example calcium \cite{Weber_2024, keller23, Pham18}, strontium \cite{Akerman15, Poulsen11}, ytterbium \cite{PhysRevA.76.052314}, barium \cite{10.1063/1.4879817}, rubidium \cite{Questionable} including Rydberg excitation \cite{BroPhysRevLett.104.010502,SafPhysRevLett.104.010503}, and cesium \cite{PhysRevX.12.011040,Mane25, Lam20}.

\subsubsection{Drawers}

Laser safety is built into the design of the drawers so that the full system operates as a Class 1 laser product. The drawers consist of a light-tight surround and an internal frame. The surround is 2-3~mm steel sheet pickled, passivated, and powder coated in RAL 9005 matte black to minimize any scattering of laser light. The drawer frame is aluminum, to save weight, and is similarly bead blasted and black anodized.

It is critical to ensure that the performance of the modules is maintained upon opening and closing of the drawers, that is without causing misalignment in the optical path. To achieve this, the drawers operate on rear-mounted cantilevered runners, mounted on the underside. This cantilever means the drawer floats, ensuring there is no change in stress on the board between the open and closed positions. This approach also provides mechanical isolation between the rack and the optical boards reducing vibrations and increasing reliability. This is further increased by a layer of rubber (3~mm thick) between optical boards and drawer bottom. Furthermore, rear-mounting allows 100\% drawer extension, increasing ease-of-use by giving the user access to the entire optical board for in-rack measurement and optimization. 

\subsubsection{Fiber Routing}

Throughout this system, laser light is transferred between modules via optical fiber. These need to be routed to and throughout the rack. This requirement introduces two important considerations. Firstly, that no fiber can be bent beyond its specified minimum bending radius (typically ten times the cable diameter \cite{FOAGuide}, here between 15~mm and 40~mm, depending on specification). Secondly, that the stress, strain, and tension on the fibers remains constant during drawer operation. Combined, this prevents damage to fibers, misalignment, slack from blocking beam paths, and allows the maximum polarization stability through polarization maintaining fibers. 

To meet these requirements, each drawer uses an energy chain for cable management. The energy chain bend radius (IGUS E16 series) is chosen to meet the minimum bend radius requirements of our optical fibers (Schäfter+Kirchhoff PMC series). The energy chains are used in combination with 3D printed fiber wheels in the rack. These are used to route the fibers whilst meeting the bend radius requirements. 3D printed cable clamps on the boards are used for strain relief of fibers and RF cables. 

\subsubsection{Board Characteristics}

The most fundamental design choice for the optical boards is the material and thickness to ensure that the board does not deform significantly under load. Increasing thickness needs to be balanced against the increased weight and cost. 

\begin{figure}[!ht]
\includegraphics[width=0.9\linewidth]{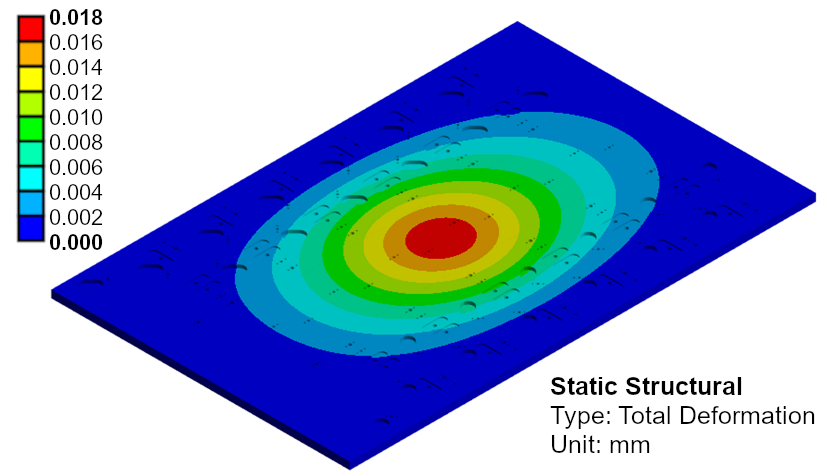}
\caption{FEA of a 12~mm aluminum custom optical board. We simulate the maximum deformation under a centrally applied 120~N load.}
\label{fig:StressFEA}
\end{figure}

The module optical boards are machined from aluminum sheets to total dimensions of 620~mm $\times$ 397~mm, 12~mm thickness, and weigh approximately 7.5~kg. This thickness was chosen following finite element analysis (FEA) of the board deformation under load. Under a centrally applied force of 120~N, the maximum vertical displacement was 0.018~mm (see Fig.~\ref{fig:StressFEA}). This is comparable to a displacement of 0.014~mm for a 3/4 inch (19.1~mm) thickness Thorlabs optical board of the same approximate dimensions (Thorlabs MBH4560/M - 600~mm $\times$ 450~mm). The weight of the components on the board total 2.0~kg and 2.8~kg for the distribution and AOM drawers, respectively. 

To ensure accurate positioning throughout, alignment sensitive components are positioned with a pair of 2~mm diameter dowel pins (H7 tolerance), fixing the position and angle of the components relative to the incident beam. Several custom optical mounts with dowel pin features are used in the modules. The design is also compatible with the alignment pin features of the Thorlabs Polaris\textregistered~line. The boards are anodized black to minimize any spurious scattered light. The anodizing process can affect the tolerance of the dowel pin holes. These critical features are therefore re-reamed after anodizing to ensure accuracy.

The height of the beam path is maintained in a plane 12.7~mm above the surface of the boards in all modules. Custom recesses and our custom mounts are used to ensure that all optical components are centered on this plane. Tight tolerances (between 0.05~mm and 0.1~mm) are used for all features that can affect alignment. Additional recesses are included to accommodate optical alignment tools (for example, Thorlabs hex key adjusters - HKTS-5/64).

\subsection{Laser stabilization system}

Stabilization of laser frequencies by locking to a fixed reference is a prerequisite for most atomic physics experiments \cite{Diode10.1063/1.1142305}. Several techniques are routinely employed to achieve this. For ion trap systems, our board designs focus on stabilization to a reference cavity \cite{Drever83}. 

Our locking system consists of custom optical boards for light delivery to a reference cavity. Each board accommodates two wavelengths, provides mode-matching optics, alignment mirrors, a dichroic combiner, and detection photodiodes in one unit. These modules, combined with the Pound-Drever-Hall (PDH) feedback loop \cite{Drever83, Poulsen11}, provide frequency stabilization for the laser light, allowing operation at the atomic transition frequencies.

\begin{figure}[!ht]
\includegraphics[width=0.9\linewidth]{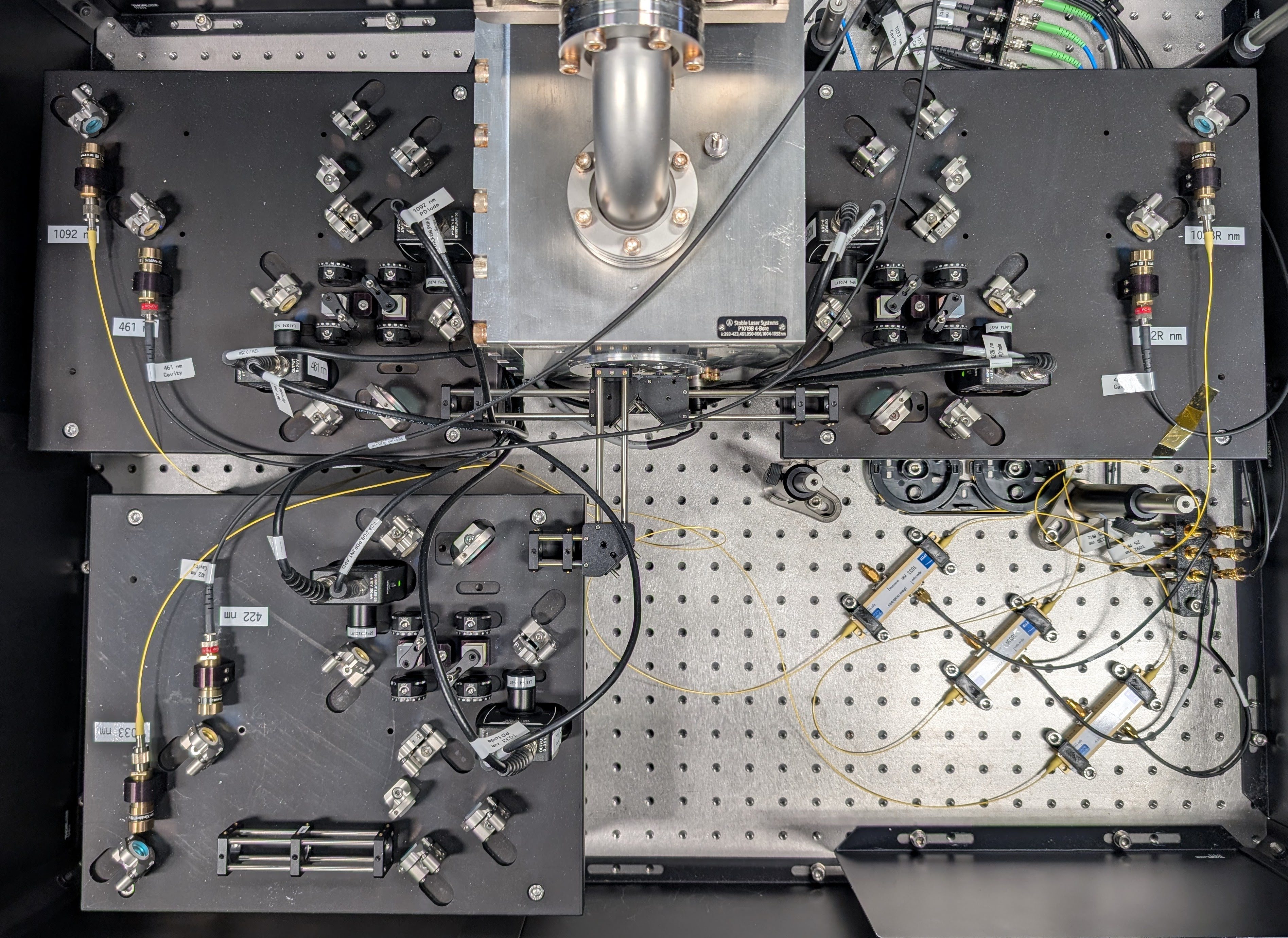}
\caption{Photo of a complete locking station. The reference cavity is surrounded by three cavity board modules (max 4). The footprint sits within a 75~cm $\times$ 90~cm passively isolated optical table. This setup stabilizes six different wavelengths independently (max 8).}
\label{fig:Cavphoto}
\end{figure}

\begin{figure}[!ht]
\includegraphics[width=8cm]{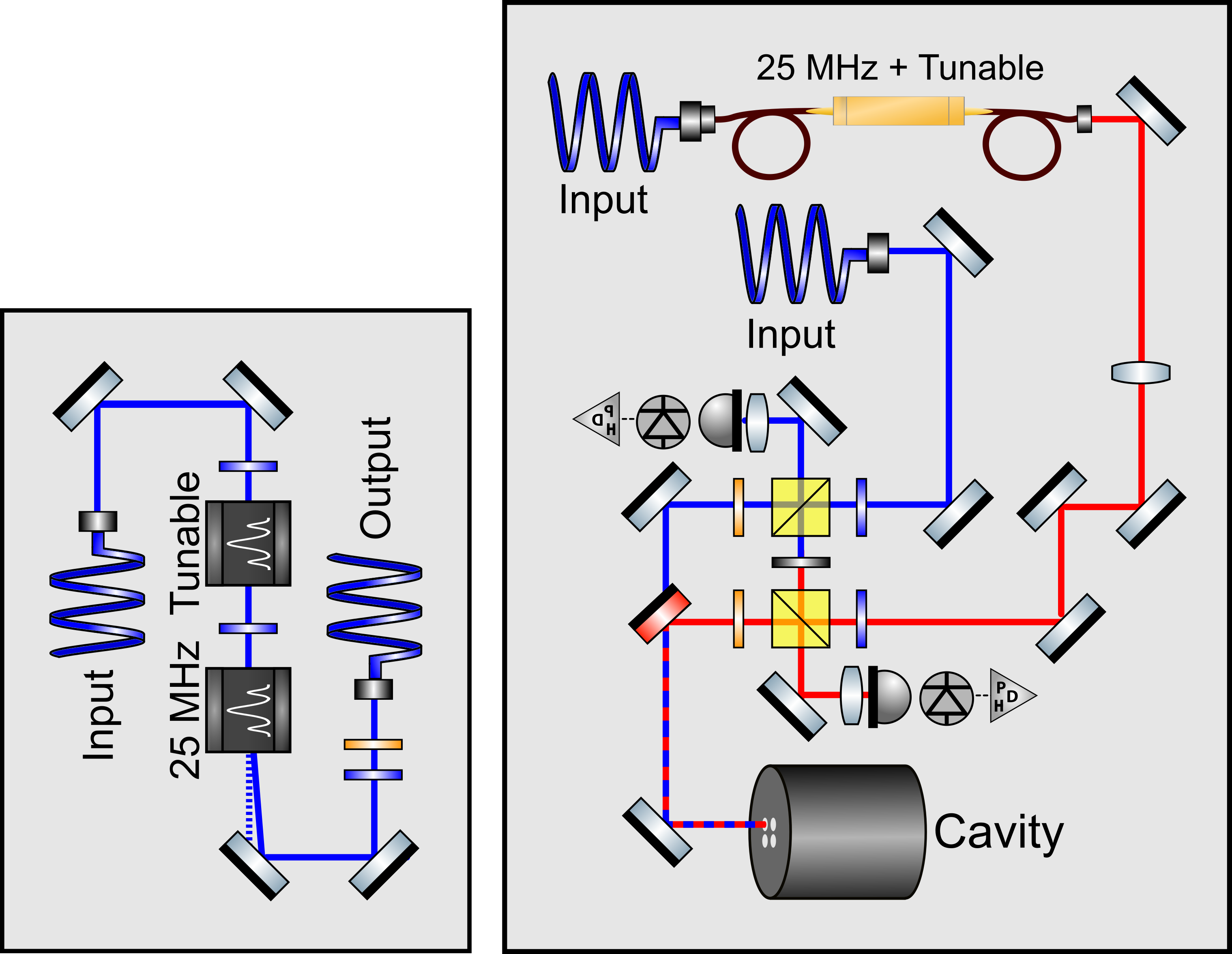}
\caption{Schematics of the optical boards used for laser stabilization. (Left) Free-space EOM module for the `blue' wavelengths. (Right) Cavity board module for delivering light to the cavity, and measuring the reflected signal. Each of these modules combines two wavelengths -- stabilizing two lasers independently (see Appendix~\ref{sec:Schematics} for the key).}
\label{fig:cavity_schematic}
\end{figure}

In a strontium ion trap setup, the laser stabilization module is composed of three of the cavity boards, stabilizing six lasers in total (see Fig. \ref{fig:Cavphoto}). For the reference cavity, we use a Stable Laser Systems (SLS) 4-bore cavity, with a cavity length of 10~cm and a Finesse within a range of $100 - 260$, wavelength dependent. The cavities are plano-concave, where the radius of curvature of the mirrors are $R_1=\infty$ and $R_2=50$~cm. Each mirror is broadband coated, allowing flexibility in wavelength. The combination of this cavity and our cavity boards allow up to eight different lasers to be stabilized simultaneously and independently. Fig. \ref{fig:cavity_schematic} shows the optical paths used to feed light into the reference cavity. 

\subsubsection{Cavity Module}

The input collimator and lenses for each beam is chosen according to the mode matching requirements discussed in Sec.~\ref{sec:modematching}. A pair of steering mirrors is used for alignment, and a PBS allows access to the reflection from the cavity. A dichroic mirror combines the two wavelengths onto a final broadband mirror which delivers both to a single cavity port. The reflected signal from the PBS is focused on an amplified photodiode (Thorlabs PDA8A2), which serves as the input signal for the feedback loop. The transmission is monitored by adding an extra photodiode or camera on the rear side of the cavity.

\subsubsection{PDH Feedback Loop}
\label{sec:floop}

\begin{figure}[!ht]
\includegraphics[width=0.9\linewidth]{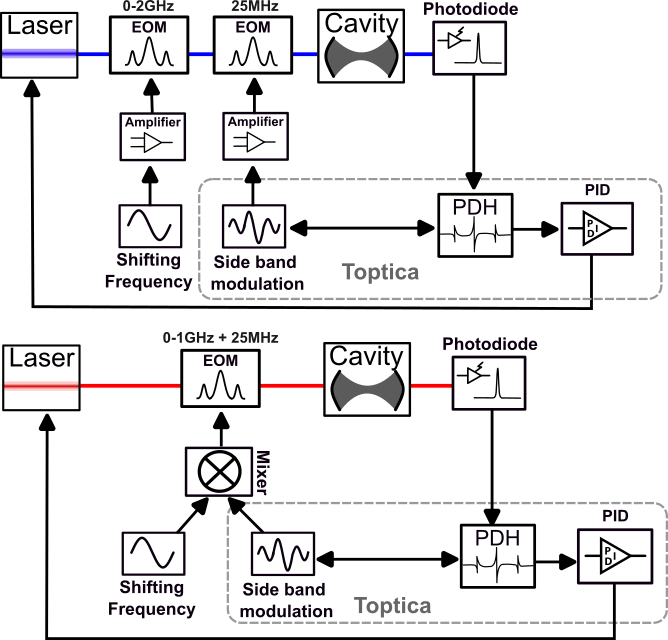}
\caption{Schematic of the locking system used to stabilize our lasers. Optical paths are represented by color traces and the electronic paths by arrows. (Top) The schematic for `blue' wavelengths. (Bottom) The schematic for `red' and `infrared' wavelengths. The PDH module (for generating the sideband modulation and error signal demodulation) and the PID feedback is provided by Toptica.}
\label{fig:feedback_loop}
\end{figure}

Our hardware is compatible with any electronic PDH implementation. We use the PDH module provided by Toptica (PDH/DLC pro). The PDH module provides an output modulation of 25~MHz or 5~MHz, which are used to generate the laser modulation for PDH sidebands. The amplitude of the sidebands and the phase of the modulation are adjusted to provide a steep slope on the central feature. In Fig. \ref{fig:feedback_loop}, a detailed scheme of the feedback loop is shown. 
The cavity has a fixed length, and as such, the resonance peaks cannot be shifted in frequency space; they are spaced by the FSR of the cavity, 1.5~GHz (Fig. \ref{fig:FSR_fullplot} Top). A second type of modulation is used to bridge the gap between the cavity resonances and the atomic transitions. This is a tuneable modulation between 50~MHz and 750~MHz applied to an EOM. The tuneable EOM creates additional sidebands which provide a PDH error signal exactly at the atomic resonance (see Fig. \ref{fig:FSR_fullplot} Bottom).

For the `red' and `infrared' EOMs, the 25~MHz and tunable modulation signals are sent to the same fiber-coupled EOM (Jenoptik Phase Modulators series), and an RF signal splitter (Mini-circuits ZX10-2-12-S+) is used to mix both inputs. For the blue wavelengths, two free-space EOMs are required, as there are no suitable fiber-coupled EOM solutions. The first EOM provides the fixed 25~MHz modulation (Qubig PM7-VIS-25). The second provides a tuneable sideband modulation up to 2~GHz (Qubig TWP2M2-VIS). These EOMs are mounted to a separate optical board connected via optical fiber to the cavity modules. The design considerations for this module include the beam sizes and polarizations required for the EOMs to reach their maximum efficiency. By using a separate EOM module, the beam size across the EOM board can be chosen to optimize transmission through the EOMs, and to be independently optimized on the cavity module for the mode-matching requirements of cavity. This EOM module design could also be used for other applications, for example to provide dual tone light for qubit preparation \cite{LucasPhysRevA.95.022337, BariumPhysRevA.81.052328}. 

\subsection{Design Requirements}

\subsubsection{Rayleigh Length Calculations}
\label{sec:AOMEff}
Gaussian beams inherently diverge as they propagate. The distance over which a beam can be considered collimated is limited to a fraction of its Rayleigh length (RL) \cite{siegman1986laser,RLRP}. Limiting the optical path length ensures the beam can be treated as collimated throughout and efficiencies in optical elements can be planned and optimized correctly. 

For the distribution modules, no element requires a specific beam size, as such there is no specific requirement on the chosen collimator. Therefore, we aim for larger beams to avoid RL constrains. A beam size of 3~mm was targeted such that the RL greatly exceeds the maximum path length in the module (maximum optical path 65~cm, minimum RL 6.5~m at 1092~nm).

For the AOM modules there is a constraint -- AOM efficiency is dependent on beam size. Two different double-pass AOMs configurations were prototyped: placing the AOM at the focus of a 1:1 telescope and placing the AOM after a beam reducing telescope. The latter was observed to have better overall efficiency and its reduced footprint allowed more modules to be included on each board. 

The path length of the AOM module is 94~cm. As a result, an RL of greater than 2~m is required, leading to input beam diameters between 1.2~mm and 1.7~mm. Using an input beam diameter of 1.3~mm for the `blues', 1.7~mm for `reds' and `infrareds' and a telescope of de-magnification $\approx 4$ (using $f_1=$100~mm, and $f_2=~-25$~mm lenses), achieves a beam diameter at the AOM of $\approx$ 325~$\upmu$m for `blues', 425~$\upmu$m for `reds' and `infrareds'. For our chosen AOMs, the maximum efficiencies are achieved for beam diameters of  $\approx$~120~-~300~$\upmu$m, wavelength dependent. This compromise is required given the trade-off between RL considerations and AOM efficiency (see Appendix~\ref{sec:AOMeff}). 
 
\begin{figure}[!ht]
\includegraphics[width=0.9\linewidth]{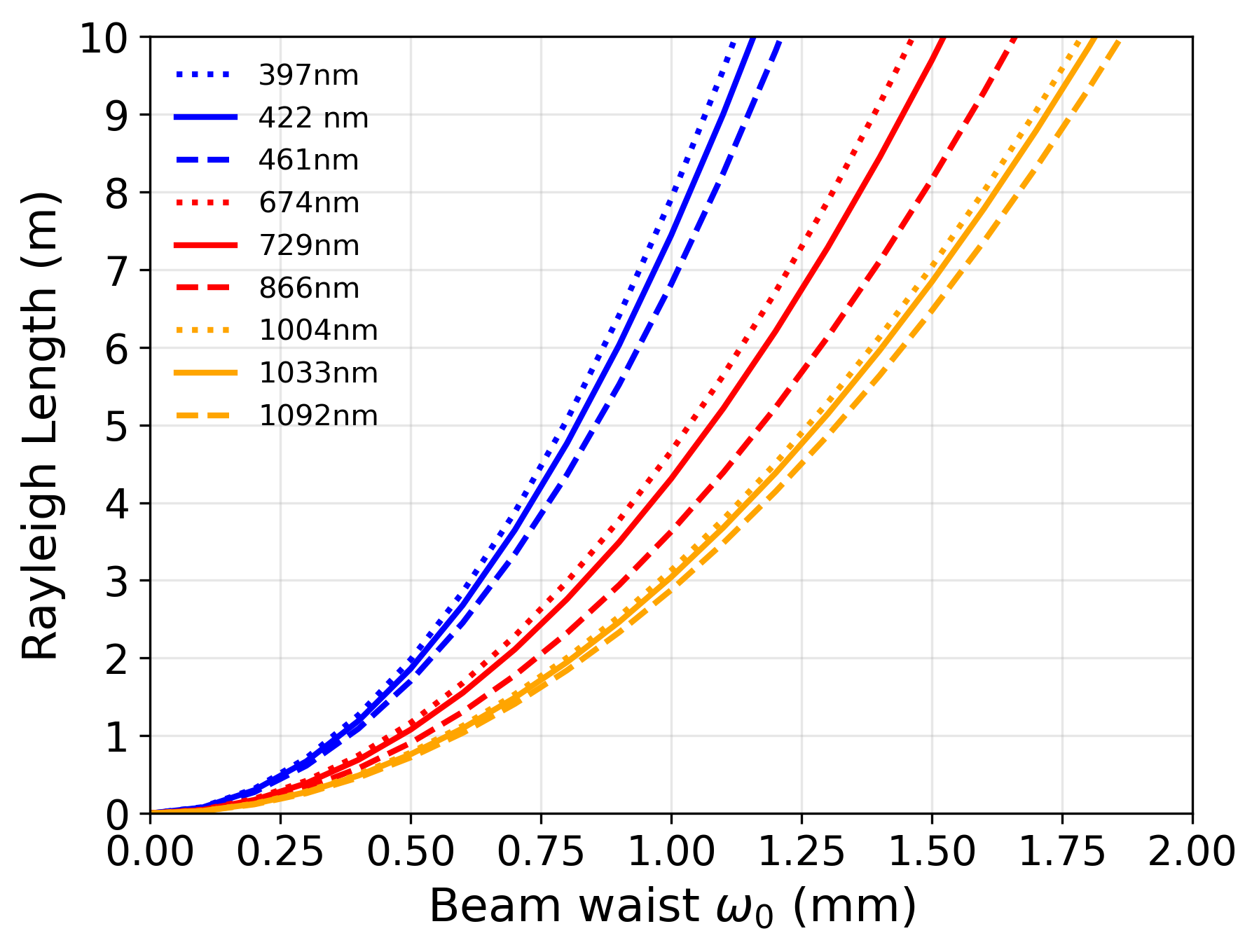}
\caption{Rayleigh length as a function of beam waist across the range of wavelengths used. We omit 423~nm and 854~nm for clarity.}
\label{fig:RL}
\end{figure}

\subsubsection{Cavity Mode Matching}
\label{sec:modematching}

Being well mode matched to the cavity is crucial for a clean, reliable locking signal. The mode waist required inside the cavity is set by the cavity length and radius of curvature of its mirrors. These quantities can then be used to calculate the mode size on the input mirror

\begin{equation}
    \omega_0 = \frac{L\lambda}{\pi}\sqrt{\frac{g_1 g_2(1-g_1 g_2)}{(g_1 +g_2-2g_1 g_2)^2}}
\end{equation}
with
\begin{equation}
    g_{i}=1 - \frac{L}{R_i}~,
\end{equation}
where $i$ denotes mirror $1$ or $2$, $R_i$ is the respective radius of curvature, $L$ is the length of the cavity, and $\lambda$ is the wavelength. The position of the waist relative to the mirrors can therefore be calculated and one can define the mode size at the mirrors using the Gaussian beam propagation relationship \cite{siegman1986laser}
\begin{equation}
w_{1,2}=w_0 \left[ 1+ \left( \frac{z_{1,2}}{z_0}\right)^2\right]^{1/2}~.
\end{equation}

Our choice of input mirror is arbitrary, since our cavity is symmetric for reflectivity, and once the spot size required is calculated, there are several ways to achieve mode matching.

Using the movable lens inside the collimator (Schäfter + Kirchhoff 60FC-SF series) to focus the beam to $w_{1,2}$ at the input mirror was preferred for mode matching. Space for multiple additional lenses is built into the board for when mode matching was not possible with the collimator alone (see Appendix~\ref{sec:MMcomp}).

\section{Module performance}
\label{sec:results}

\subsection{Distribution}

Distribution modules are characterized by the fiber coupling efficiency of the output collimator and the polarisation extinction ratio (PER), defined as the ratio between two orthogonal polarizations and their respective transmission $\text{ER}=P_{\text{min}}/P_{\text{max}}$ \cite{goldstein2003polarized}, when correcting for the degree of unpolarized light. Losses in all other optical components are considered negligible. 

\subsubsection{Fiber Coupling}

Fiber coupling is achieved using a pair of steering mirrors and the integrated adjustable lenses on the collimators. The fiber coupling efficiency was measured across 8 wavelengths, and 53 fiber in-couplings, and is -- on average -- above $60\%$ for blue wavelengths, above $70\%$ for red wavelengths, and above $75\%$ for infrared wavelengths. This behavior is expected, given the fiber mode field diameter (MFD), defined as $\text{MDF}\approx\frac{2\lambda}{\text{NA} \pi}$, is proportional to wavelength ($\approx 4.0~\upmu$m for blue fibers compared to $\approx 8.3~\upmu$m for the infrared fibers) making angular and positional alignment more sensitive at shorter wavelengths. NA is the numerical aperture of the fibre. Additionally, the attenuation coefficient is higher for shorter wavelengths, and approaches a minima at 1550~nm \cite{FundOpt8}.

\begin{figure}[!ht]
\includegraphics[width=0.9\linewidth]{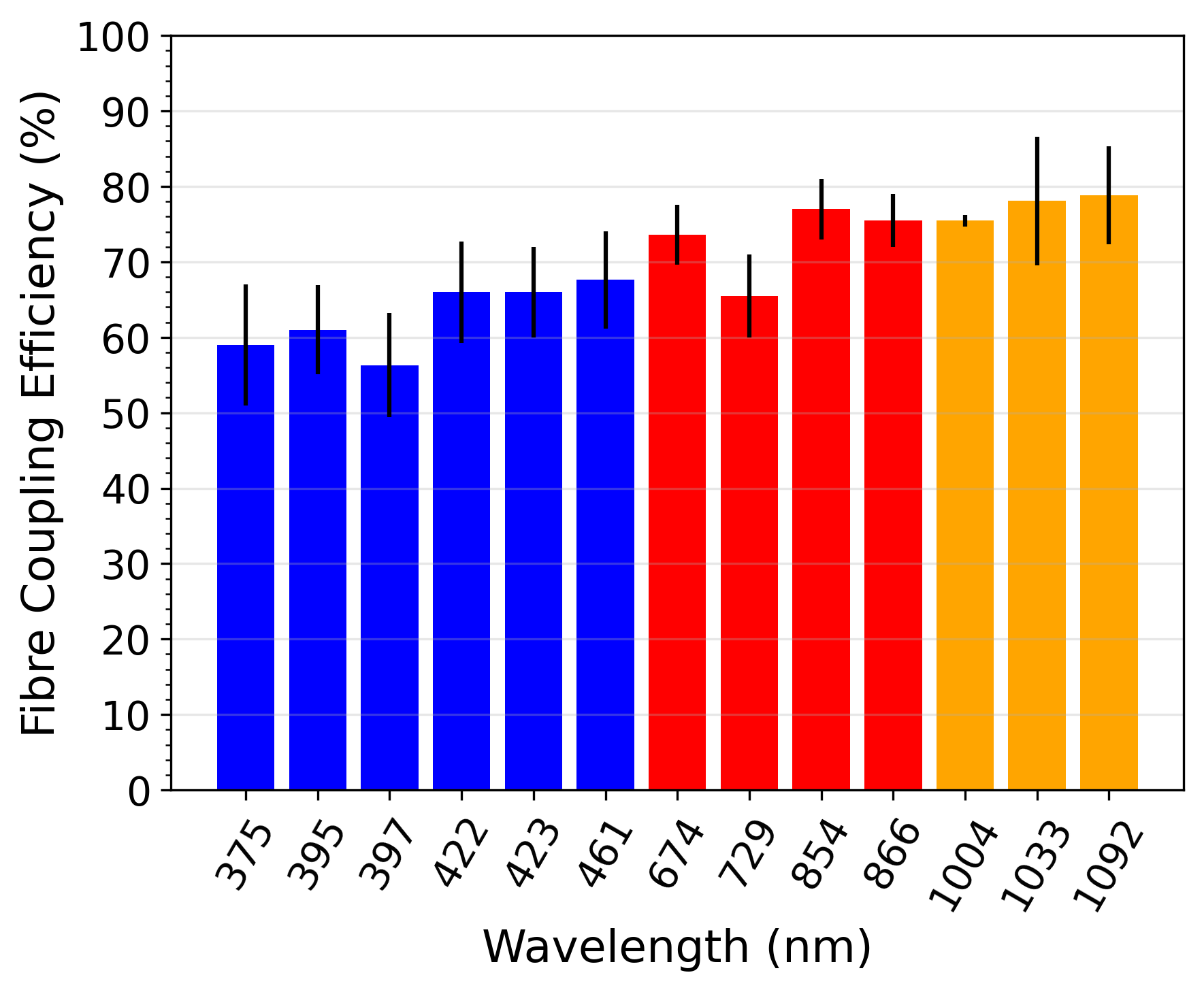}
\caption{Fiber coupling efficiencies. Measured over 53 fiber in-couplings on the distribution modules -- averaged by wavelength.}
\label{fig:FibreEff}
\end{figure}

\subsubsection{Polarization}

All fibers are polarization maintaining (PM), specified to have an PER of 40~dB (Schäfter + Kirchhoff, PMC series). To achieve this PER, the polarization must be linear and aligned to the fast (or slow) axis of the fiber \cite{fibrecouplingAxis}. Full control over the polarization is achieved with the combination of a half- and quarter-waveplate. We minimize the polarization displacement on a Poicaré sphere \cite{goldstein2003polarized,LabPoinc}, when stressing the fiber, as measured by a polarimeter (Thorlabs PAX1000). A PER of 40~dB or greater was achieved for all fibers. 

\subsubsection{AOM Modules}

In the AOM modules, the obtained fiber coupling efficiencies and PER values were consistent with the distribution board values. The combined AOM double-pass and output fiber coupling efficiency for a range of wavelengths is presented in Fig.~\ref{fig:FibreEff}. We achieve $\approx 70\%$ efficiency for both the first and second pass through the AOM (consistent across all wavelengths). This is consistent with the stated manufacturer efficiencies (see Appendix~\ref{sec:AOMeff}). Alignment and efficiency measurements are performed at the peak efficiency of the AOM. 

\begin{figure}[!ht]
\includegraphics[width=0.9\linewidth]{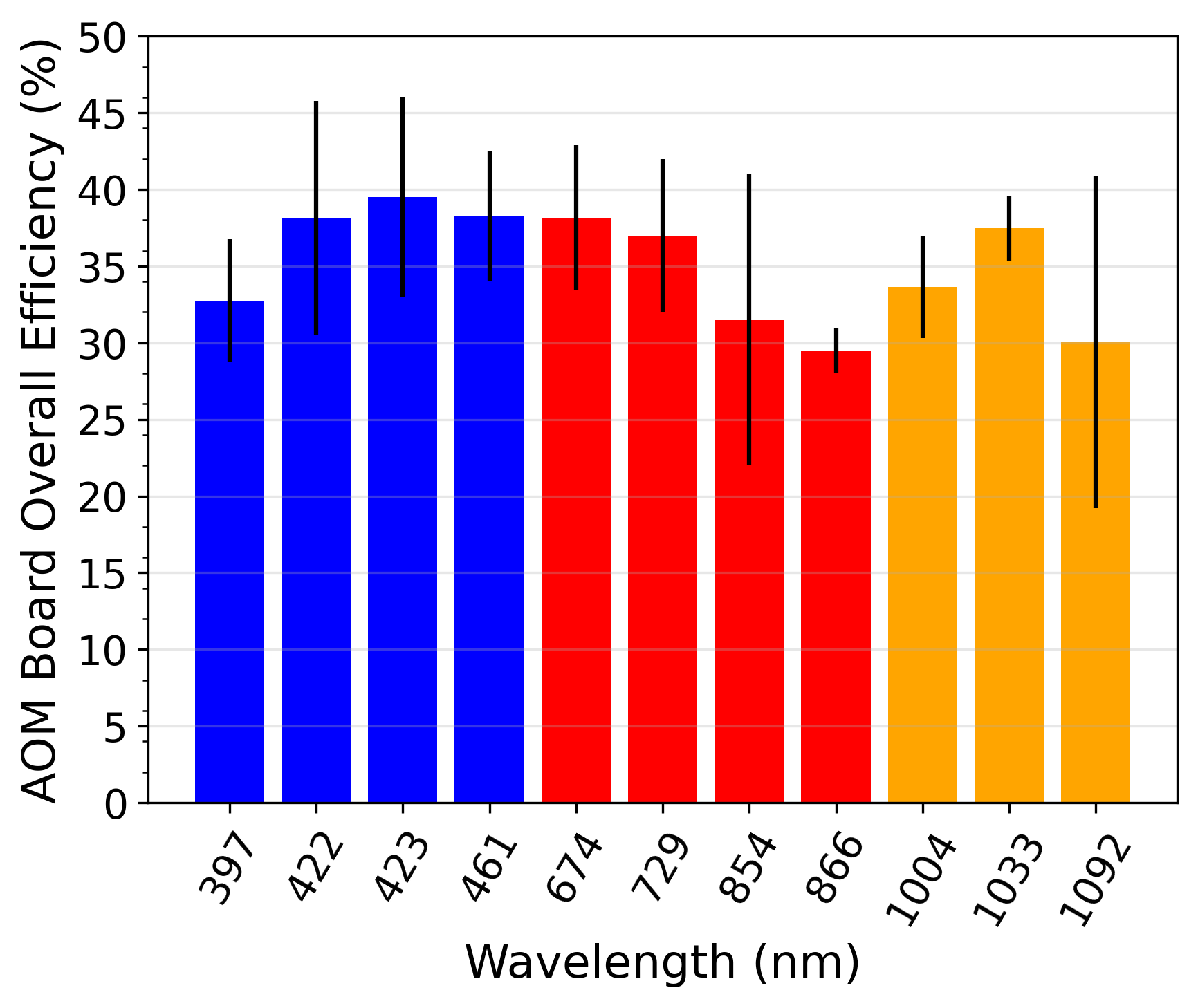}
\caption{Overall efficiencies of the AOM modules -- combining double-pass AOM and fiber coupling efficiency.}
\label{fig:AOMeff}
\end{figure}

\begin{figure}[!ht]
\includegraphics[width=0.9\linewidth]{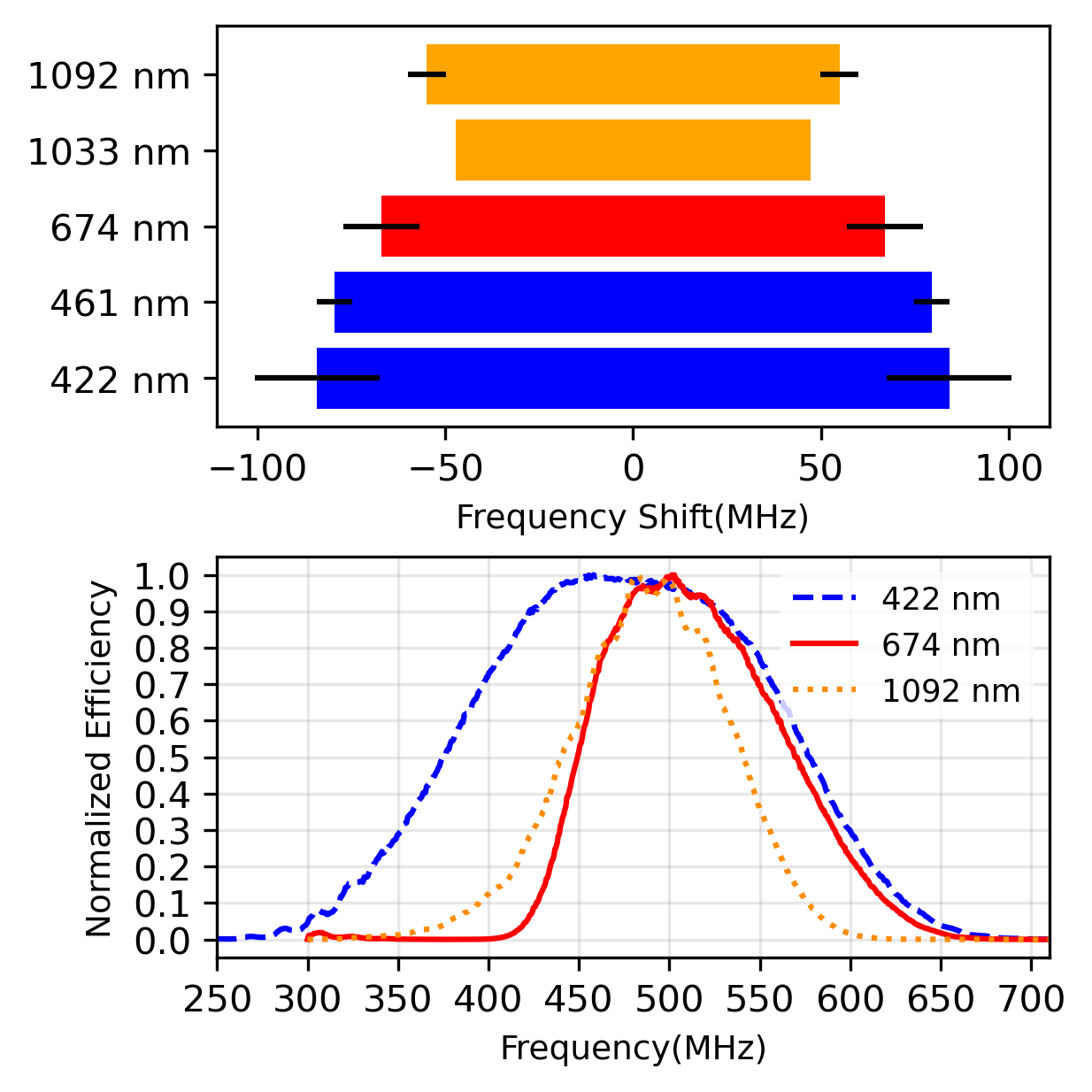}
\caption{Effective bandwidths of the AOM modules. (Top) The total optical frequency shift of the effective bandwidth at a 250~MHz central frequency. (Bottom) The normalized efficiency curves for selected wavelengths, used to identify the effective bandwidths (one for each AOM model used).}
\label{fig:BWplot}
\end{figure}

The bandwidth of an AOM gives the maximum frequency shift that can be applied before the efficiency is reduced by half. Here we define the `effective bandwidth' as the double-pass bandwidth measured after the fiber coupling, that is the frequency shift applied for the total efficiency of module to reduce by half. To extract the effective bandwidth, we perform a Gaussian fit to the measured data (see Fig.~\ref{fig:BWplot} (Bottom)), and take the full width at half maximum (FWHM). 

The double-pass configuration allows frequency shifts whilst minimizing the displacement of the diffracted beam. However, we observe the effective bandwidth to be smaller than the nominal bandwidth specified for the AOMs. The results shown in Fig.~\ref{fig:BWplot} (Top) indicate the usable frequency ranges of the AOM modules. The total frequency shift is above $\pm 50$~MHz for all AOMs. 

Many standard techniques in atomic physics, for example, optical pumping and state preparation, are performed near resonance, as pumping rates are inversely proportional to the detuning from atomic transition \cite{suter1997physics}. For laser cooling, Doppler cooling uses frequency detunings of the order of the natural linewidth of the atomic transition $\Gamma$ \cite{Metcalf1999Laser,suter1997physics}, while sideband cooling use detuning equivalent to the ion motional frequency $\omega_T$ \cite{Sidebandcool1,QuantumDynam}. As both $\Gamma$ and $\omega_T$\ are smaller than $50$~MHz -- for typical atomic species -- the module's effective bandwidth is sufficient to cover the required frequency scanning. For techniques that require larger detunings, for example, polarization gradient cooling \cite{Li_2022,Dalibard:89}, additional range can be obtained by scanning the laser lock-point via the tunable EOM in the locking system.

Polarization noise in optical fibers is typically the limiting factor in laser power stability. The fibers used have -- to the best of the authors' knowledge -- the highest commercially available PER at the time of writing. To characterize the power stability of the entire system, we measured the power over 15 minutes at the experimental outputs. We calculate the coefficient of variation (CV) defined as: $\text{CV}=\sigma / \mu$, where $\sigma$ is the standard deviation and $\mu$ is the average of the measured signal. For one implementation of our system the measured values for each  wavelength were: $\text{CV}_{422}=0.69\%$, $\text{CV}_{461}=0.79\%$, $\text{CV}_{1004}=0.90\%$, $\text{CV}_{1033}=0.68\%$, $\text{CV}_{1092}=0.15\%$. Long-term stability of the systems has been observed over several months with no need for re-alignment. Active power stabilization can be easily implemented with the addition of an appropriate pick-off optical element and RF drive feedback to the AOM.

\subsection{Locking System Performance}
\label{sec:Locking System Performance}

\subsubsection{EOM Module Performance}

The overall transmission efficiency through the EOM module is 20\% - 30\%. This is 10\% - 15\% lower than what we would expect given the nominal efficiencies of the EOMs. The efficiency could be improved by adding a further pair of mirrors between the two EOMs, allowing them to be optimized independently. However, as the total power required for the locking system is low ($\approx200\upmu$W), the overall power loss is not a concern and we prioritize compactness of the module. 

\subsubsection{Cavity Alignment}

\begin{figure}[!ht]
\includegraphics[width=0.9\linewidth]{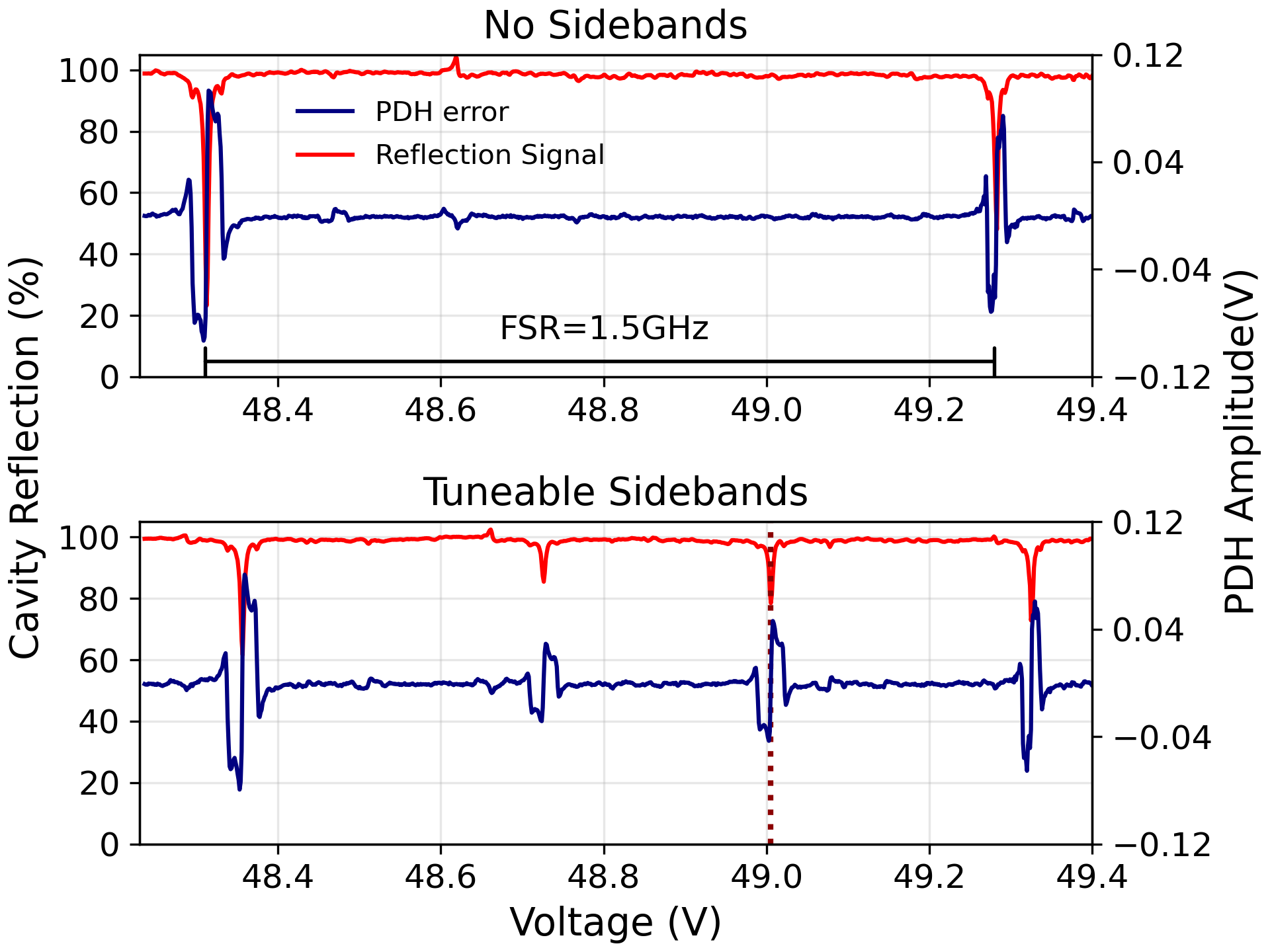}
\caption{(Top) Reflection signal (red) and PDH error signal (blue) as a function of scanning the piezo voltage in the laser. The bracket indicates the spacing between two natural occurring cavity resonances, i.e. the FSR (422~nm laser in this example). (Bottom) Same signal with the `tuneable sidebands' applied to match the atomic frequency. The dotted line indicates the frequency of the atomic resonance. }
\label{fig:FSR_fullplot}
\end{figure}

Using irises in detachable cage-mounted alignment jigs, the beams are coarsely aligned before being positioned on the table. Cage mounts position the board with respect to the cavity enclosure. A beam-splitter with a CCD camera and a photodiode are mounted on the back of the cavity to monitor transmission. With this minimal alignment, an initial signal (higher order Hermite-Gaussian modes \cite{siegman1986laser}) is seen on the camera. Further alignment using the steering mirrors, is used to observe the fundamental Hermite-Gaussian $\text{TEM}_{00}$ mode \cite{siegman1986laser}. Using the transmission photodiode, one can then observe the full-spectral range (FSR) by scanning the laser. Further beam-walking and adjustment of the collimator focusing lens is used to increase the coupling to the $\text{TEM}_{00}$, and minimize other modes. Mode-matching above $70\%$ on the $\text{TEM}_{00}$ mode is achieved for all wavelengths (for example see Fig. \ref{fig:FSR_fullplot}, Top). 

\subsubsection{Locking Performance}

\begin{figure}[!ht]
\includegraphics[width=0.9\linewidth]{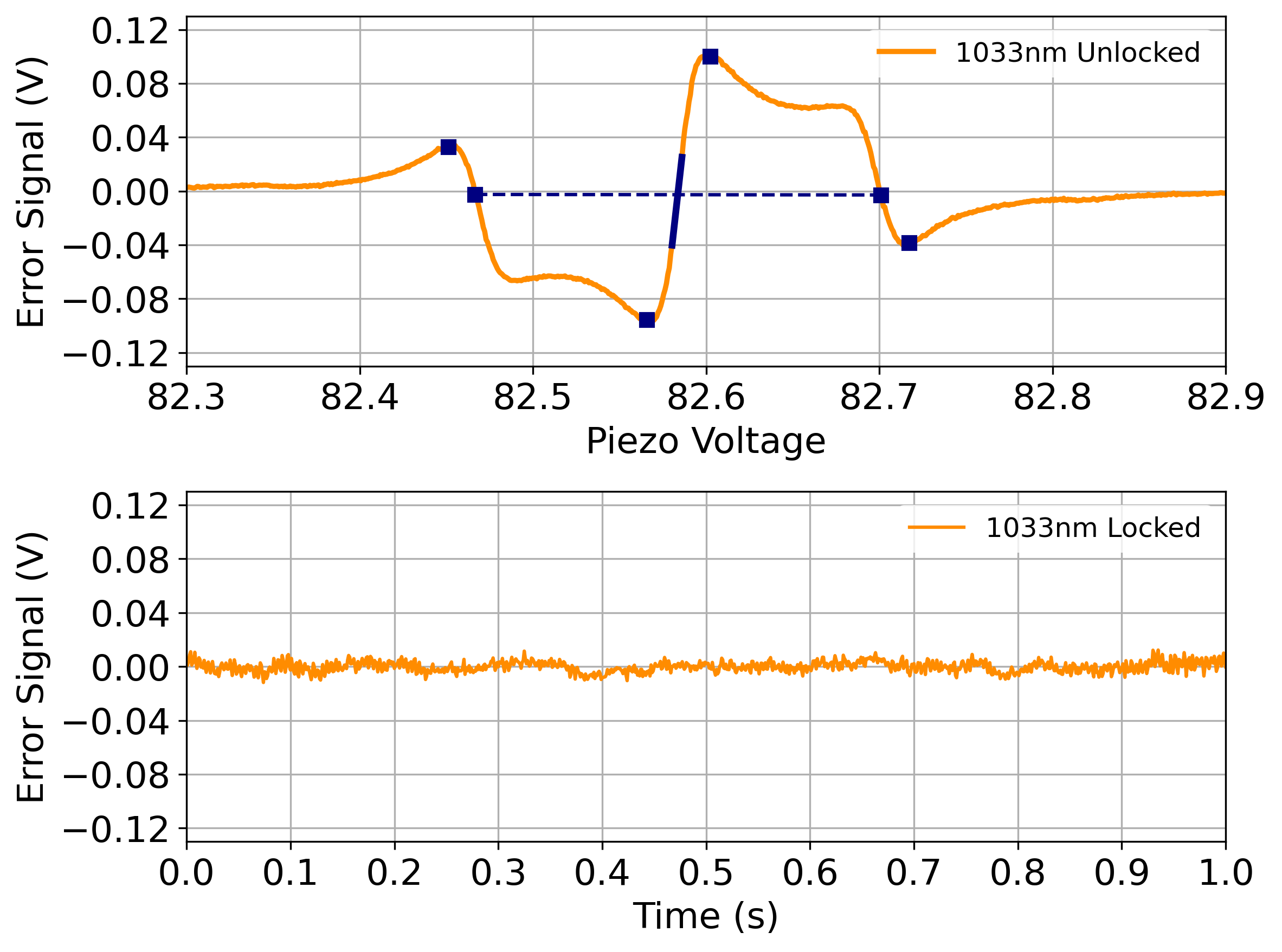}
\caption{Linewidth estimation. (Top) Example PDH error signal. The dots identify the inflection points in the signal and are used to fit the central slope (solid blue line) and the 25~MHz sideband separation (dashed line) (1033~nm laser in this example). (Bottom) Error signal with the laser locked. The standard deviation of this signal is used to estimate the linewidth.}
\label{fig:PDH_locking}
\end{figure}

All wavelengths are stable to within 2~MHz over a period of an hour. This corresponds to the absolute precision of the wavemeter used (HighFinesse WS8-10). The linewidth is a more precise measurement of stability. An estimate of linewidth is obtained by measuring the standard deviation of the locked signal over 1~s (Fig.~\ref{fig:PDH_locking}, Bottom) and converting it into frequency. By fitting a linear function to the center slope of the PDH error signal (blue in Fig.~\ref{fig:PDH_locking}, Top) we extract a relationship between error signal amplitude deviation and piezo voltage. The variation in piezo voltage can be directly mapped into frequency variation, using either the FSR or the separation of the 25~MHz sidebands as a reference. Combining the above, we obtain the conversion of error signal amplitude to a frequency deviation that is used to estimate the linewidth. The results are presented in Table~\ref{tab:example1}. The estimated linewidth of all the lasers was below 500~kHz. This performance is sufficient for the majority of wavelengths in atomic physics applications \cite{clock0,Clock1,clock2,clock3}.    

\begin{table}[!ht]
\centering
\begin{tabular}{|c|c|}
\hline
Wavelength (nm) & Linewidth (kHz) \\ \hline
397 & 201.4 \\ \hline
422 & 297.4 \\ \hline
423 & 319.7 \\ \hline
461 & 182.5 \\ \hline
854 & 37.6  \\ \hline
866 & 100.6 \\ \hline
1004 & 102.2 \\ \hline
1033 & 311.2 \\ \hline
1092 & 329.7 \\ \hline
\end{tabular}
\caption{Estimated linewidth across a range of wavelengths.}
\label{tab:example1}
\end{table}

\section{Outlook}
\label{sec:Outlook}

The laser systems developed here represent a modular and high performance system to control AMO experiments. The designs cover applications such as laser cooling and trapping, optical pumping and state preparation, and coherent control of atoms. Our approach offers fully tested designs using trusted components to significantly reduce the setup time required for AMO experiments. The modular approach also offers easy extensibility and modifications. 

Additional modules under consideration include alternative approaches to laser stabilization for cold atoms, such as saturated absorption spectroscopy \cite{Raveninbook} or modulation transfer spectroscopy \cite{MartinezdeEscobar:15}, a fiber noise cancellation module \cite{fibrecan1, Fibrecan2} (for high-fidelity coherent operations requiring low phase noise \cite{noise1,noise2,noise3,noise4}), and dichroic boards for combining several broad wavelengths using achromatic reflective collimators.

\section{Conclusion}
\label{sec:Conclusion}

We have detailed the concept, development, and characterization of a modular, high performance laser system for AMO physics. Our approach shifts the implementation of a complete laser system from a time-consuming setup task towards the procurement of a stand-alone product. 

\begin{figure}[t]
\vspace{5mm}
\includegraphics[width=0.9\linewidth]{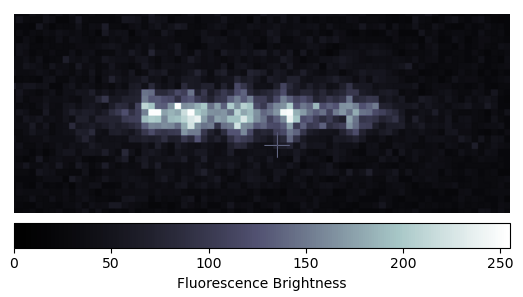}
\caption{Chain of 5 $^{88}\text{Sr}^{+}$ ions trapped in a micro-fabricated 3D-trap with our modular laser system.}
\label{fig:Ion_chain}
\end{figure}

Our design focuses on performance, safety, compactness, robustness, and cost. We have built three instances of laser system, demonstrating stable operation and ion trapping. Fig.~\ref{fig:Ion_chain} shows a chain of 5 strontium ions in a 3D microfabricated trap \cite{NPL1, NPL2} using the modular system reported here. The overall footprint of the system is a 19-inch rack, together with a 75~cm~$\times$~90~cm optical table for laser stabilization. The system is portable and has been transported between two labs 160~km apart. Minimal re-alignment was required to recover the specified performance after transport.

The system allows one to distribute, stabilize, and control frequency and amplitude of up to 4 experimental outputs for each laser. The performance was characterized by some key metrics: frequency stability with linewidths $\leq500$~kHz, polarization extinction ratios $\geq40$~dB, overall power delivery (from laser source to experimental output) $\geq21\%$, free-running power stability $\text{CV}<1\%$, and frequency tuning $>\pm 50$~MHz. In addition, the systems are cost effective. A single custom optical board costs $<$~£450: cheaper than a comparably sized standard commercial breadboard. A complete double-pass AOM module costs $\approx$~£4400: significantly cheaper than commercially available alternatives, with assembly and optimization of a board of four AOM modules taking around one day.

As quantum technologies move towards product-like deployment, solutions like the one demonstrated here become an integral part of scaling systems. Furthermore, a modular approach means that parts can be upgraded or replaced with minimal disruption to the overall platform.

\leavevmode\par
\section{Design Availability and Licensing}

The designs presented in this work (and future extensions) are available for licensing, and we welcome inquiries from potential partners. Please contact the authors or STFC Business \& Innovation for further information (\href{mailto:innovations@stfc.ac.uk}{innovations@stfc.ac.uk}).

\section{Acknowledgments}

The authors acknowledge the significant contribution of Alex Jones from the STFC Technology Department to the mechanical design of the system. The strontium ion trap system (a vacuum-packaged ion microtrap with electronic interfacing, optomechanical beam delivery, optical imaging, and magnetic field coils) was developed and constructed at the National Physical Laboratory by William Broe, Alastair Sinclair, Guido Wilpers and Stephanie Webster. That work was funded by the Knowledge Assets Grant Fund of the UK Government Office for Technology Transfer. Yaron Bernstein and Elizabeth Bain from the STFC Business and Innovation Directorate supported the licensing of the designs.

\bibliographystyle{apsrev4-2}
\bibliography{references}

\clearpage
\appendix
%\documentclass[reprint,amsmath,amssymb,aps,pra,floatfix]{revtex4-2}
%\usepackage[utf8]{inputenc}
%\usepackage{graphicx}
%\usepackage{xcolor}
%\usepackage{braket}
%\usepackage{physics}
%\usepackage[title]{appendix}
%\usepackage[colorlinks]{hyperref}

%\begin{document}

%\title{Supplementary material -- Design and implementation of a modular laser system for AMO experiments}

%\author{Klara Theophilo, Scott J Thomas, Georgina Croft, Yashna N D Lekhai, Alexander Owens, Daisy R H Smith, Silpa Muralidharan and Cameron Deans}
%\email{cameron.deans@stfc.ac.uk}
%\affiliation{National Quantum Computing Centre, Rutherford Appleton Laboratory, OX11 0QX, UK}
%\date{\today}

%\maketitle

\section{Laser sources}
\label{sec:Laser_sources}
All of our lasers were supplied by Toptica. Calcium system lasers:
\begin{itemize}
\item External-Cavity Diode Lasers (ECDL) (MDL pro) -- T-rack integrated: 397~nm, 423~nm, 854~nm, 866~nm
\item ECDL Tapered Amplified Laser (MTA pro) -- T-rack integrated: 729~nm
\item iBeam smart: 375~nm
\end{itemize}
Strontium system lasers:
\begin{itemize}
\item ECDL Laser (MDL pro) -- T-rack integrated: 422~nm, 461~nm, 1004~nm, 1033~nm, 1092~nm
\item Modular Frequency-Doubled Amplified Laser System (DLC TA-MSHG PRO) -- T-rack integrated: 674~nm
\item iBeam smart: 395~nm
\end{itemize}

\section{Key for Schematics}
\label{sec:Schematics}

\begin{figure}[!ht]
\includegraphics[width=0.5\linewidth]{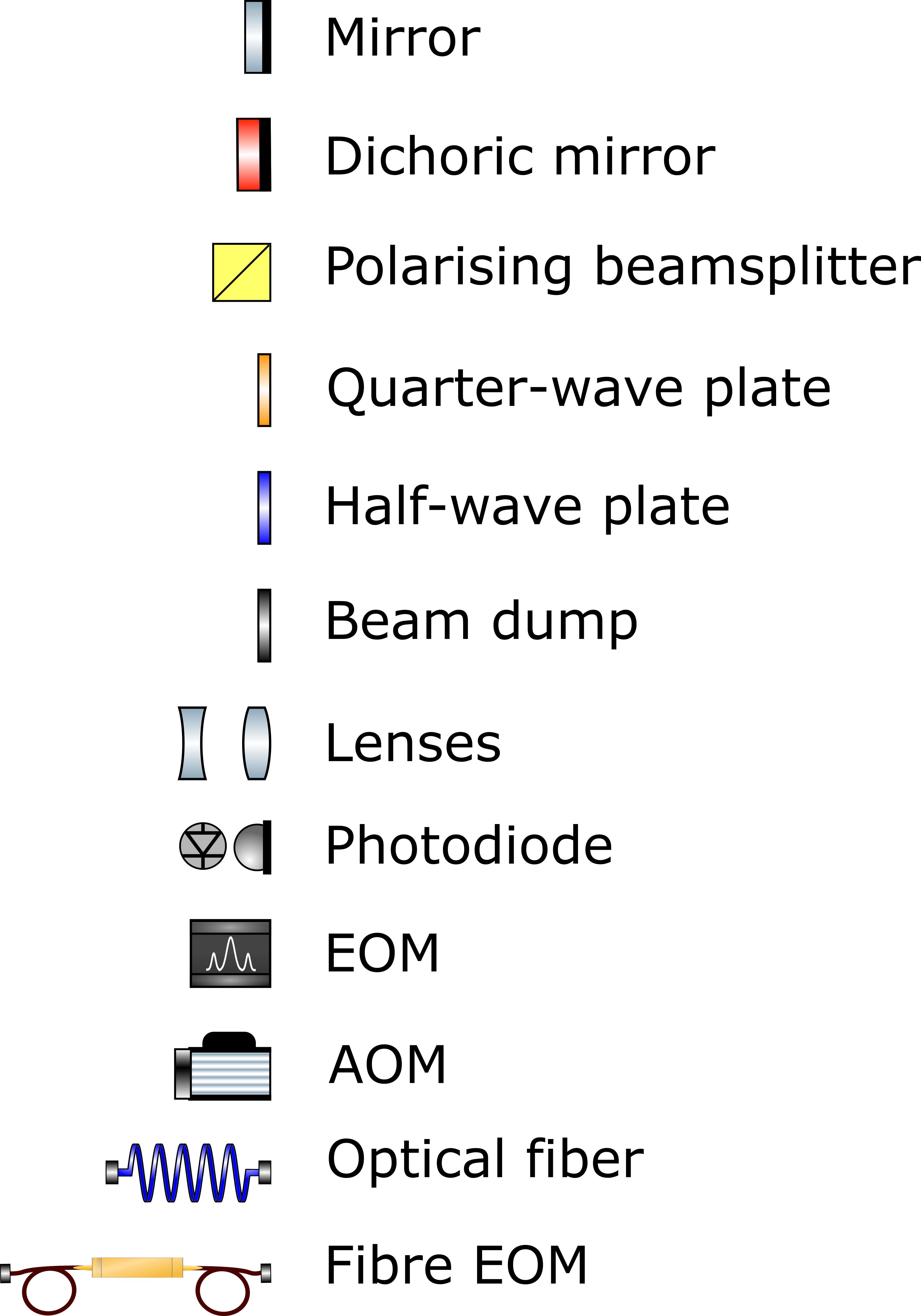}
\caption{Key for schematic diagrams.}
\label{fig:key}
\end{figure}

Fig.~\ref{fig:key} is the key for the optical schematic diagrams in the main text.

\section{Comparison of AOM performance and specifications}
\label{sec:AOMeff}

The quoted peak efficiency and transmission for each model of AOM used is included in Table~\ref{tab:AOMspecs}. The recommended beam diameter for the `red' and `infrared' AOMs (AA Opto-Electronics) is between $200-300~\mu$m and $120~\mu$m for the `blue' AOMs (Isomet). Our beam diameters are $\approx450~\mu$m and $\approx325~\mu$m, respectively. These values result from balancing several competing constraints: Rayleigh length, optical path length, minimizing footprint, maximum efficiency, and the products available. In total, our double-pass efficiency is only around $8\%$ below the specified maximum (comparable across the four models).

\begin{table*}[t]
\centering
\caption{AOM specifications for the four models of AOM used.}
\label{tab:AOMspecs}
\begin{tabular}{|c|c|c|c|c|}
\hline
AOM Model & Wavelength (nm) & Efficiency & Transmission & Bandwidth (MHz) \\
\hline
\hline
MT250-B100A0.5-VIS (AA Opto-Electronics) & 687 and 674 & $>85\%$ & $>95\%$ & 100 \\
\hline
MT250-B100A0.2-800 (AA Opto-Electronics) & 866 and 854 & $>60\%$ & $>95\%$ & 100 \\
\hline
MT250-B100A0.5-1064 (AA Opto-Electronics) & 1004, 1033, and 1092 & $>70\%$ & $>95\%$ & 100 \\
\hline
M1250-T250L-0.45 (Isomet) & 397, 422, 423, and 461 & $>85\%$ & $>95\%$ & 100 \\
\hline
\end{tabular}
\end{table*}

\begin{figure}[!ht]
\includegraphics[width=0.9\linewidth]{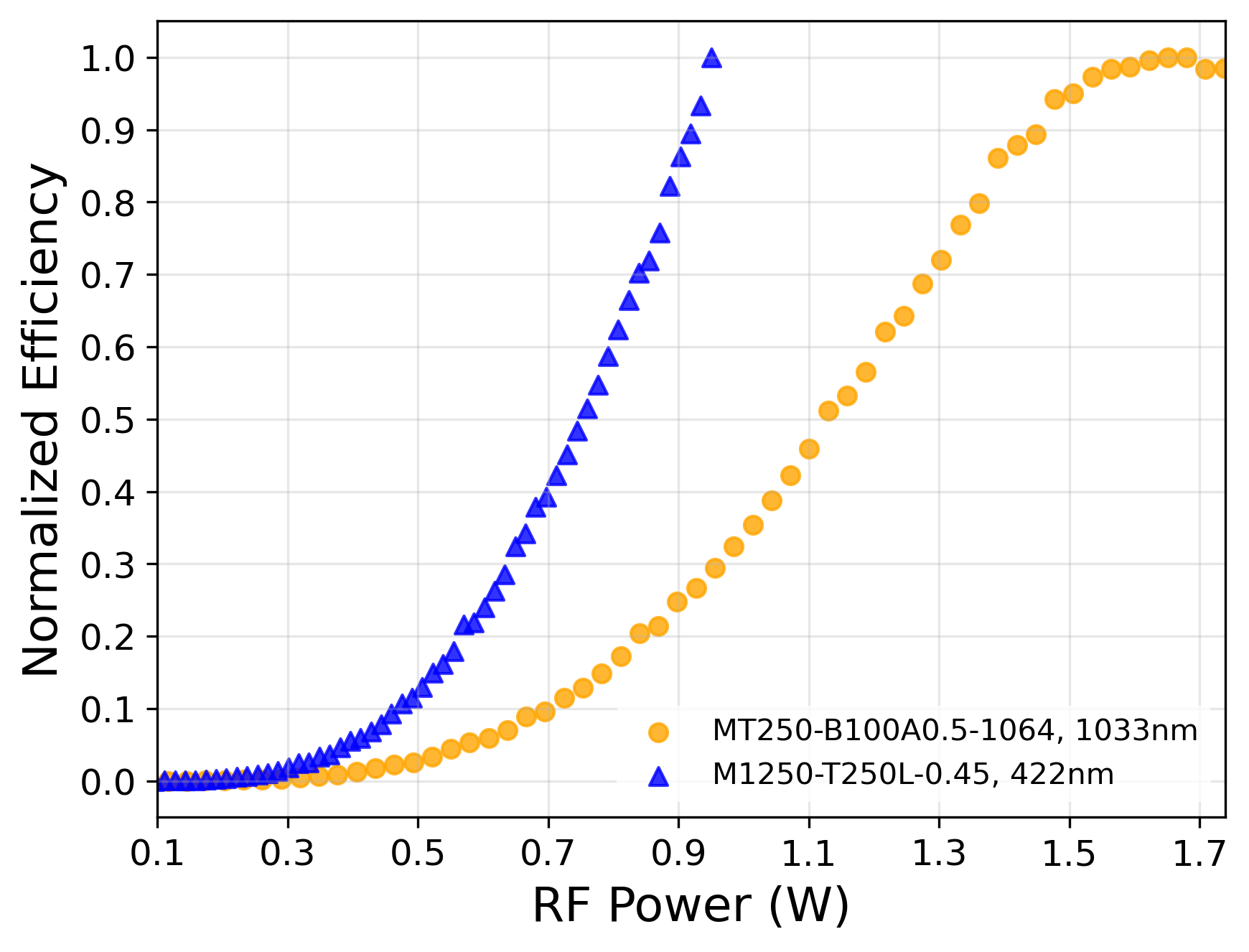}
\caption{AOM normalized efficiency against the RF input. The RF power is measured at input of the AOM, to account for losses in the coaxial cables.}
\label{fig:RF_eff}
\end{figure}

Typical AOM responses of the two manufacturers against RF driving power can be seen in Fig.~\ref{fig:RF_eff}. The expected saturation of the diffraction efficiency is not observed for the `blue' AOMs. The manufacturer stated that this is not the expected behavior. However, this response is consistent across all 26 AOM modules from the three implementations of the system. We limit the RF driving power to the specified maximum to avoid damage.

\section{AOM rising times}
\label{sec:RisingTime}

\begin{figure}[!ht]
\includegraphics[width=0.9\linewidth]{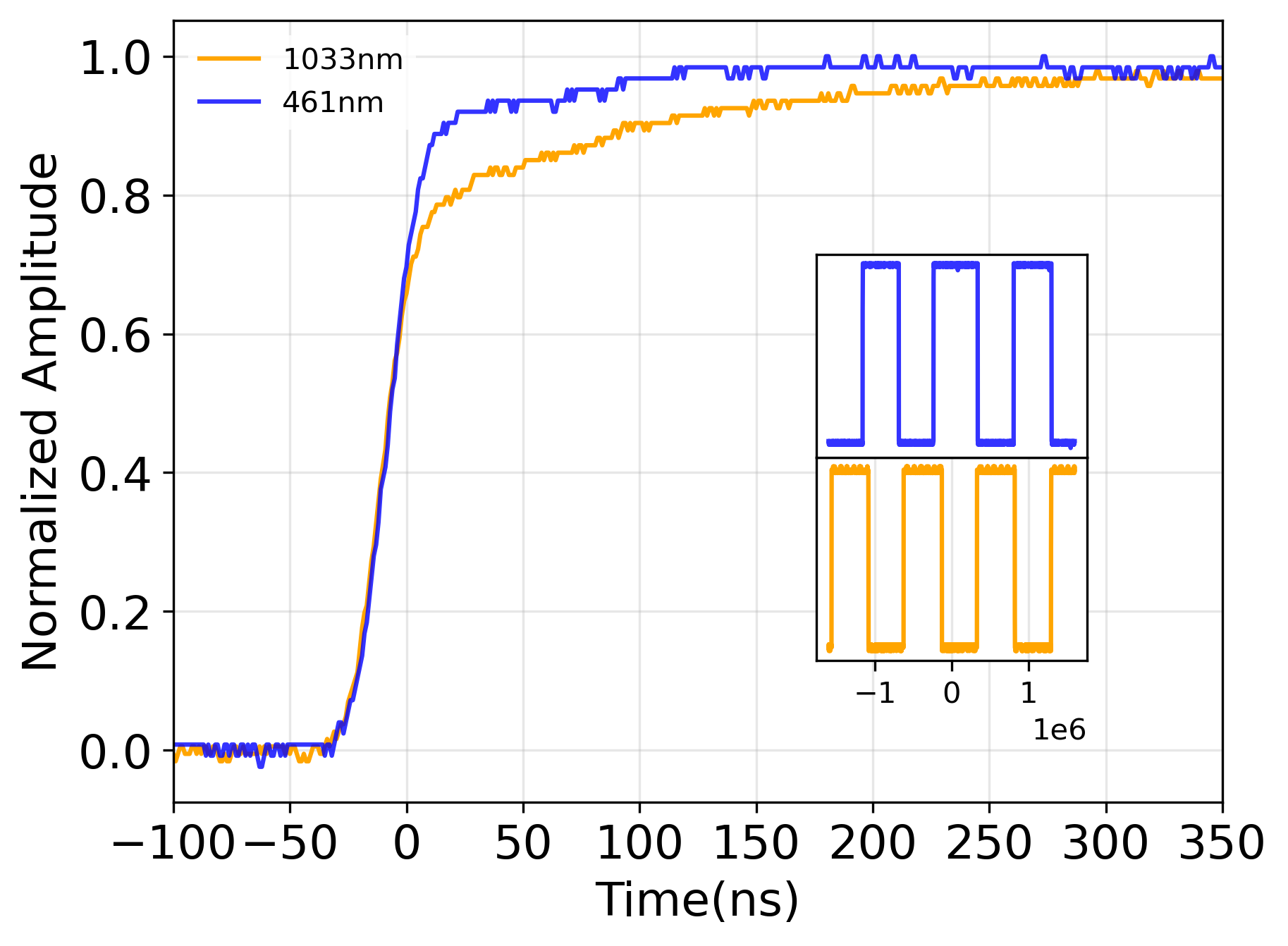}
\caption{Example rise times of two AOM models. One from the `blue' wavelengths (Isomet M1250-T250L-0.45) and one from the `infra-red' wavelengths (MT250-B100A0.5-1064). Inset: Optical response with a 1~MHz square wave applied to the AOM.}
\label{fig:Pulse_shape}
\end{figure}

In our setup, the AOM rise time is $\approx 20$~ns. The AA Opto-Electronics AOMs also display a settling time of a few hundreds of nanoseconds -- see Fig.~\ref{fig:Pulse_shape}. Two modifications to the system should be considered if fast pulsing (sub-microsecond) is required. Firstly, the temperature of the AOM should be kept constant to avoid the increased settling time. This can be achieved by driving the AOM with constant RF power (i.e. adding second RF tone outside the bandwidth when the AOM is off) or by combining the AOM pulsing with a mechanical shutter (to minimize the time that the RF drive is switched off). Secondly, the AOM rise time is inversely proportional to the incident beam diameter. The diameter, and consequently the rise time, can be decreased by focusing the light onto the crystal -- e.g. placing the AOM at the center of a 1:1 telescope. This approach comes at the cost of reduced overall efficiency and increased footprint.

\section{Complementary equations for mode matching}
\label{sec:MMcomp}

In the main text we introduced the main equations needed to estimate mode matching. In addition, the position of the waist and the position of the mirrors relative to the waist are needed. The position of the waist $z_0$ is
\begin{equation}
z_{0}=\frac{d(R_1-d)(R_2-d)(R_1+R_2-d)}{(R_1+R_2-2d)^2}.
\end{equation}
The position of the mirrors $z_{1,2}$ is
\begin{equation}
z_{1,2}=\frac{-d(R_{2,1}-d)}{R_1+R_2-2d}
\end{equation}
where $d$ is the separation between two mirrors, and $R_i$ are the radius of curvature of each mirror $i=1,2$.

Additional lenses may be required to optimize the spot-size for mode matching. To calculate the focal length $f$ required to achieve the necessary diameter $S_d=2w_{1,2}$ at each mirror we use
\begin{equation}
    f = \frac{\pi D S_d}{4 \lambda M^2}~,
    \label{eq:spotsize}
\end{equation}
where $D$ is the input collimated beam diameter, $\lambda$ is the wavelength, and $M^2$ is the beam mode parameter (assumed to be $1$ for a collimated beam). There is a further requirement that a lens of focal length $f$ must be placed at that distance away from the cavity mirror along the optical path. Our cavity boards use cage mounted lenses to allow a 10~cm variation in the position of additional lenses to account for this.

%\end{document}

\end{document}